\documentclass[apj]{emulateapj}
 
\slugcomment{To appear in {\it{The Astrophysical Journal}}}
\shortauthors{Gonzalez et al.}
\shorttitle{A $z=2.79$ Lensed LIRG Behind the Bullet Cluster}

\newcommand \firstp {Paper I}
\newcommand \firstpaper {(Paper I)}
\newcommand{\kms}{~km~s$^{-1}$}
\newcommand{\spitzer}{\textit{Spitzer}}

\newcommand{\herschel}{\textit{Herschel}}
\newcommand{\hst}{\textit{HST}}

\newcommand{\irac}{IRAC}
\newcommand{\irs}{IRS}

\newcommand{\wfc}{WFC3}
\newcommand{\acs}{ACS}
\newcommand{\kcorrect}{\texttt{kcorrect}}
\newcommand{\galfit}{\texttt{GALFIT}}
\newcommand{\irsclean}{\texttt{IRSCLEAN}}
\newcommand{\spice}{\texttt{SPICE}}
\newcommand{\multidrizzle}{\texttt{MultiDrizzle}}
\newcommand{\hyperz}{\texttt{HyperZ}}
\newcommand{\ergflux}{~erg~s$^{-1}$~cm$^{-2}$}
\newcommand{\micr}{$\mu$m}
\newcommand{\target}{{the lensed galaxy}}
\newcommand{\invmab}{$(\mu_{AB}/100)^{-1}$}

\begin{document}
\title{Spectroscopic Confirmation of a $z=2.79$ Multiply Imaged
  Luminous Infrared Galaxy Behind the Bullet Cluster}
  
\author{Anthony H. Gonzalez} 
\affil{ Department of Astronomy, University of Florida, Gainesville, FL 32611-2055}
\email{anthony@astro.ufl.edu}

\author{Casey Papovich}
\affil{Department of Physics, Texas A\&M University, College Station, TX 77843-4242} 

\author{Maru\v{s}a Brada\v{c}}
\affil{Department of Physics, University of California at Davis, One Shields Avenue, Davis, CA 95616}
 
\author{Christine Jones}
\affil{Harvard-Smithsonian Center for Astrophysics, 60 Garden St., Cambridge, MA 02138}

\begin{abstract}
  We report spectroscopic confirmation and high-resolution infrared
  imaging of a $z=2.79$ triply-imaged galaxy behind the Bullet
  Cluster. This source, a \spitzer-selected luminous infrared galaxy
  (LIRG), is confirmed via polycyclic aromatic hydrocarbon (PAH)
  features using the \spitzer~Infrared Spectrograph (\irs) and
  resolved with \hst~\wfc~imaging. In this galaxy, which with a 
  stellar mass $M_*\approx4\times10^9$ M$_\sun$ is one of the two
  least massive ones studied with \irs\ at $z>2$, we also detect
  $H_2\;S(4)$ and $H_2\;S(5)$ pure rotational lines (at 3.1$\sigma$
  and 2.1$\sigma$) -- the first detection of these molecular hydrogen
  lines in a high-redshift galaxy. From the molecular hydrogen lines
  we infer an excitation temperature $T=377^{+68}_{-84}$ K. The
  detection of these lines indicates that the warm molecular gas mass is
  $6^{+36}_{-4}$\% of the stellar mass and implies the likely existence of a substantial
  reservoir of cold molecular gas in the galaxy.
  Future spectral observations at
  longer wavelengths with facilities like the \herschel\ {\it Space
    Observatory}, the Large Millimeter Telescope, and the Atacama
  Pathfinder EXperiment (APEX) thus hold the promise of precisely
  determining the  
  total molecular gas mass.
  Given the redshift, and using refined astrometric positions from the
  high resolution imaging, we also update the magnification estimate
  and derived fundamental physical properties of this system. The
  previously published values for $L_{IR}$, star formation rate, and
  dust temperature are confirmed modulo the revised magnification;
  however we find that PAH emission is roughly a factor of five
  stronger than would be predicted by the relations between $L_{IR}$
  and $L_{PAH}$ reported for SMGs and starbursts in \citet{pope2008}.

\end{abstract}

\keywords{galaxies: evolution, starburst --- gravitational lensing --
galaxies: clusters: general}

\section{Introduction}
\label{sec:intro}

An important legacy of the \spitzer~{\it Space Telescope} is that it
enabled the first detection of dust emission from a large number of
luminous infrared galaxies at $z>2$
\citep[e.g.,][]{papovich2006,perez2005}, demonstrating that these
galaxies dominate the massive galaxy population at high redshift. The
implication is that at $z\sim 2-3$ massive galaxies are rapidly
assembling their stars and growing supermassive black holes.  For the
most infrared-luminous systems, spectra from the \spitzer\ Infrared Spectrograph
(IRS) have provided important insights into the physical processes
driving $L_{IR}$ \citep[e.g.,][]{houck2005,yan2007}. \citet{pope2008}
used \irs\ to demonstrate that the mid-IR properties of ultraluminous
infrared galaxies (ULIRGs, $L_{IR}>10^{12}$ L$_\sun$) and
submillimeter galaxies (SMGs) are distinct, with star-formation
dominating the infrared emission for typical SMGs.  It has been
postulated by this group and others that such differences indicate
different evolutionary stages, as the dominant source of $L_{IR}$
transitions from star formation to AGN emission.

Existing studies like \citet{pope2008} clearly demonstrate that \irs~
spectroscopy provides a clean means of disentangling the AGN and star
formation contributions. A limitation to such work though is that
current samples only probe the bright end of the luminosity function
at high-z \citep{dey2008,dye2008}, and \irs~observations are practical
for only a small subset of the most luminous sources. Observations of
luminous infrared galaxies (LIRGS, $10^{11}<L_{IR}<10^{12}$ L$_\sun$),
the dominant class of IR galaxies at these redshifts
\citep[$\sim10\times$ higher density than ULIRGS]{caputi2007}, have
simply not been feasible.  Gravitational lensing provides a means of
circumventing this limitation, however, and several recent programs
exploit strong lensing to study the properties of lensed galaxies at
mid-IR and submillimeter wavelengths \citep[][and references
therein]{knudsen2008,rigby2008}.

In this paper we utilize a similar strategy to study a highly
magnified luminous infrared galaxy that lies behind the Bullet Cluster
\citep{markevitch2002}.  First identified in \spitzer~\irac~imaging
\citep{bradac2006}, this galaxy has been shown to be a triply imaged
optical drop-out \citep[][hereafter \firstp]{gonzalez2009} that is
extremely luminous at submillimeter wavelengths
\citep[$\sim 0.11$ Jy at 500$\mu$m][]{wilson2008,blast2009,johansson2010,rex2010}. Photometric redshift
estimates based upon both the stellar SED and the submillimeter
emission place the galaxy at $z\sim2.7-2.9$
\citep{wilson2008,gonzalez2009,blast2009}, while lensing constraints
imply magnification factors of $10-50$ for each of the three images
\firstpaper.  This galaxy therefore presents a unique opportunity to
study in detail the properties of a member of the LIRG
population during the peak epoch of star formation and galaxy assembly
when these galaxies may be responsible for more than half the total infrared emission \citep{rodighiero2010}.

We present in this paper the results of targeted observations of this
lensed galaxy with \spitzer~\irs~spectroscopy and \hst\ Wide Field
Camera 3 (\wfc) infrared imaging. We combine these observations with
both refined positions from the \wfc\ imaging and published results
from \citet{wilson2008}, \firstp, and \citet{blast2009} to derive
updated physical parameters for this galaxy. We also constrain the
total contribution of a central active galactic nuclei (AGN) to the
total mid-infrared emission. The data are presented in
\S\ref{sec:data}, while the analysis and derived physical properties
are presented in \S\ref{sec:analysis}. We summarize our results in
\S\ref{sec:conclusions}. Throughout the paper we assume $H_0=70$ \kms,
$\Omega_M=0.27$, and $\Omega_\Lambda=0.73$, consistent with the
five-year WMAP cosmological analysis \citep{komatsu2009}.

\section{Data}
\label{sec:data}
\subsection{\spitzer~\irs}

We observed the galaxy on 2009 April 14--15 (PI Gonzalez, program ID 496)  using the Infrared
Spectrograph \citep[IRS]{Houck2004} on board \spitzer\ during a single
\spitzer\ Astronomical Observation Request to minimize alignment
uncertainty.  We targeted the source using the \irs\ LL1 module, which
provides a spectrum in the wavelength interval $19-38$ \micr.  This
module was selected based upon the existing photometric redshifts,
which indicated that the prominent polycyclic aromatic hydrocarbon
(PAH) emission features should lie in this wavelength interval.

\begin{figure}
\plotone{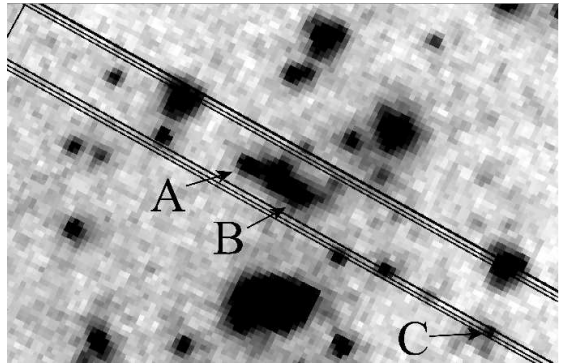}
\caption{Shown here are the dithered \irs~slit positions overlaid on
  the 8$\mu$m image of the region. The three images of the LIRG are
  indicated with arrows and labelled A, B, and C. With this slit
  location we recover the combined spectrum of images A and B, but are
  not able to recover a spectrum for image C. Note that there is an
  elliptical galaxy that lies directly between images A and B in the
  8$\mu$m image. By 24$\mu m$ this elliptical contributes negligible
  flux in comparison to the two images of the LIRG \firstpaper. The
  field of view is 65$\arcsec\times$50$\arcsec$. North is up and east
  is to the left.\label{fig:image}}
\end{figure}

We centered the long slit on image B of the galaxy (see Figure
\ref{fig:image}).  We used high--accuracy peak-up centroiding of stars
from 2MASS \citep{Skrutskie1997} with the IRS red filter to minimize
deviations from the absolute telescope pointing.  We did not restrict
the spacecraft orientation for our data in order to maximize the
likelihood that our observations would be executed prior to the end of
the \spitzer\ cryogenic mission.  Fortuitously, the slit orientation
for our observations, 240.9\degr\ east of north, placed both images A
and B of the source within the slit.  Therefore our IRS observation
contains the combined flux from the two brightest images of the
galaxy.  We were unable to extract a spectrum for image C, which lay
on the very edge of the slit (see Figure \ref{fig:image}).  We obtained our
\irs\ observations in mapping mode with four 20$\arcsec$ steps along
the IRS slit.  The total exposure time for the LL1 spectroscopy is $4
\times 3.6$ ks, where each map position consisted of 30 cycles with
120s ramp duration.

We reduced the IRS data using software provided by the \spitzer\
Science Center (SSC),\footnote{http://ssc.spitzer.caltech.edu} and
custom scripts \citep[see ][]{papovich2009} based on techniques
discussed in \citet{teplitz2007}.  We started with the S18.7.0
\spitzer\ \irs\ pipeline data, which produced basic calibrated data
(BCD) files.  The IRS/LL arrays accumulate latent charge during long
exposures \citep{teplitz2007}.  We measured and subtracted this latent
charge by fitting a first--order polynominal to the mean counts per
BCD as a function of time.

We next identified and cleaned known bad and hot pixels using the SSC
task
\irsclean\footnote{http://ssc.spitzer.caltech.edu/dataanalysistools/tools/irsclean/}
with the known warm--pixel mask for our IRS campaign.  We also
identified other ``rogue'' and ``warm'' pixels as those with
abnormally high variance, and used \irsclean\ to interpolate over
these pixels.

We constructed a sky spectrum for \target\ at each of the four slit
positions by combining the BCDs of the three other slit positions.  To
create the sky image, we took the median of the stack of each pixel
after rejecting outliers using a sigma--clipping algorithm.  We
performed this process iteratively, masking out the location of
\target\ during subsequent iterations (\target\ is the only source we
identify in the 2D spectrum).  We then subtracted the sky frame from
each BCD and coadded the BCDs at each slit position.  As a last step,
we reran \irsclean\ on the combined images for each slit position to
clean any remaining hot pixels.  These steps produced four 2D
spectroscopic images for \target, one at each of the four slit
positions.

We extracted 1D spectra at each slit position using the \spitzer/IRS
custom extraction (\spice)
software\footnote{http://ssc.spitzer.caltech.edu/postbcd/spice.html}.
Following \citet{teplitz2007} we used a narrower extraction window
than the default to minimize the noise contribution from the
background.  We used a window that had a width of $\approx$3.5 pixels
at 27~$\mu$m and scaled with the wavelength.  We estimated an aperture
correction for this narrower aperture by comparing the spectra
extracted using both the narrow and optimal extraction regions for the
standard star HD~163466 taken in the same campaign as our science
data.  However, because the IRS observations of this star were taken
in starring mode, and because our science target is likely extended
along the IRS slit, this aperture correction is unlikely to be
perfect.  The total flux density of the IRS spectrum integrated with
the MIPS 24$\mu$m passband we derived is $S(24\mu\mathrm{m})=0.925$
mJy, only slightly lower than $S(24\mu\mathrm{m})=0.965\pm0.028$ mJy
coadded from images A and B from photometry in the MIPS 24~\micron\
data in \firstp.  Therefore the aperture correction we applied seems
valid.  Nevertheless, we multiplied the IRS spectrum by a factor of
1.04 to normalize to the MIPS data, thereby accounting for any
remaining aperture effects and flux calibration issues.
As noted in \citet{gonzalez2009}, the spectral energy distributions of the
cluster elliptical and star that lie near the galaxy fall rapidly with 
wavelength and contribute negligibly at 24$\mu$m. We therefore require no
correction for these sources in the \irs~spectral extraction.

\begin{figure}
\plotone{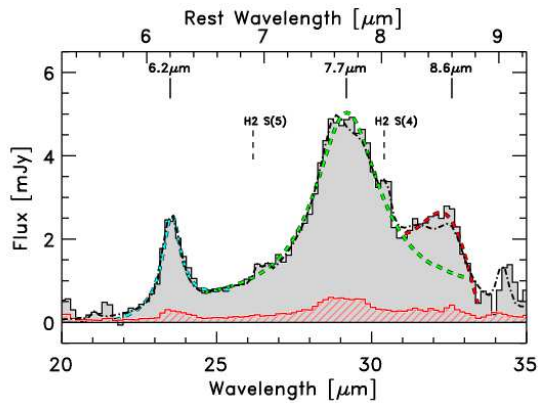}
\caption{The \irs~spectrum of the galaxy, taken in the long-low
  mode. The data are presented as the shaded histogram, with the
  uncertainties shown by the hashed histogram. The dot-dashed line
  indicates the best fit from PAHFIT, while the dashed curves
  correspond to fits to the individual PAH lines using Drude profiles
  and the formulae from \citet{smith2007}. The vertical marks above
  the spectrum denote all spectral features robustly detected in our
  analysis. \label{fig:spectrum}}
\end{figure}

To evaluate the errors in our spectra, we also extracted a sky
spectrum using the same \spice\ parameters at each slit position
offset from \target.  We use these sky spectra to estimate the error
derived from the science spectra.  Our tests showed that the errors
derived from the sky were consistent with those propagated directly
through the data reduction, and we use the latter in subsequent
analysis.  We then combined the four one--dimensional spectra as a
weighted mean, using the weights derived from the error spectra.  The
estimate of the variance on the combined science spectrum is the
inverse sum of the weights.  Figure \ref{fig:spectrum} shows the
resultant one--dimensional science spectrum.

\subsection{\hst\ Imaging}

We obtained imaging with the \hst\ Wide Field Camera 3
\citep[\wfc][]{kimble2008} on 19-20 November 2009 (Cycle 17, proposal
11099, PI Brada\v{c}), and with the Advanced Camera for Surveys
\citep[\acs][]{Ford2003} on 2006 October 12-13 (Cycle 15, proposal
10863, PI Gonzalez) and 2004 October 21 (Cycle 13, proposal 10200, PI
Jones). The new \wfc\ data consist of two overlapping \wfc/IR
pointings in F110W and F160W, with total integration times at the
location of the lensed LIRG (where the \wfc\ pointings overlap) of
13ks and 14ks, respectively. The \acs\ data, which are described in
\firstp, include F606W, F775W, and F850LP imaging.

All images were processed using standard calibration files and stacked
using custom software from the HAGGLeS project (Marshall et al. 2010,
in preparation) that is based upon
\multidrizzle~\citep{koekemoer2002}. To register the F850LP, F110W,
and F160W images with the astrometric accuracy needed for this
analysis (in particular since the standard distortion model we used
did not give sufficient accuracy to simply align overlapping
pointings), we determine the offsets among the images by extracting
high S/N objects in the individual, distortion corrected exposures. We
use SExtractor \citep{Bertin1996} and the IRAF routine geomap to
identify the objects and calculate the residual shifts and rotation of
individual exposures, which were then fed back into Multidrizzle. We
use {\it square} as the final drizzling kernel and an output pixel
scale of 0.1 arcsec; this is smaller than the original pixel scale of
the WFC3/IR, but larger than ACS CCD, allowing us to exploit the
dithering of the observations and improve the sampling of the PSF.

\begin{deluxetable}{ll}
\tabletypesize{\scriptsize}
\tablecaption{Observed Fluxes and Magnitudes\tablenotemark{a}}
\tablewidth{0pt}
\tablehead{
\colhead{Quantity} & \colhead{Value}
}
\startdata
f(6.2\micron) & $1.4\pm0.2\times10^{-14}$ \ergflux\ \\
f(7.7\micron) & $6.3\pm1.2\times10^{-14}$ \ergflux\ \\
f(8.6\micron) & $5.2\pm3.3\times10^{-14}$ \ergflux\ \\
f($H_2 S(4)$) & $5.8\pm1.9\times10^{-15}$ \ergflux\ \\
f($H_2 S(5)$) & $2.5\pm1.2\times10^{-15}$ \ergflux\ \\
f(7.7\micron)/f(6.2\micron) & $4.5\pm1.1$ \\
m$_{F160W}$ &	$23.80\pm0.1$ (AB) \\
\enddata
\label{tab:observed}
\tablenotetext{a}{All quoted values are for the combination of images A and B.}
\end{deluxetable}

\section{Analysis and Results}
\label{sec:analysis}

\subsection{PAH Features and Redshift}

The \irs~spectrum for the galaxy is presented in Figure
\ref{fig:spectrum}. The 6.2$\mu$m, 7.7$\mu$m, and 8.6$\mu$m PAH
complexes are detected in the spectrum. We also tentatively detect the
$H_2 0-0 S(4)$ line, which is discussed separately in \S
\ref{sec:molecular}.  We use Drude profiles, as in \citet{smith2007},
to model each spectral feature and use a power law to fit the
underlying continuum. We derive the redshift from the two strongest
PAH features (6.2\micr\ and 7.7\micr), obtaining $z=2.791\pm0.007$.
This redshift confirms the photometric redshifts in the literature
\citep[$z\sim2.7-2.9$;][]{wilson2008,gonzalez2009,blast2009}.

The derived fluxes for the PAH features are listed in Table
\ref{tab:observed}. The flux ratio for the two highest signal-to-noise
lines, $f(7.7\mu\mathrm{m})/f(6.2\mu\mathrm{m})= 4.5\pm1.1$, can be
compared with results from \citet{pope2008} for submillimeter galaxies
(SMGs). The star-formation dominated SMGs in the Pope et al. sample
($z\sim1-2.5$) have flux ratios in the range of 1.4 to 3.5 for these
PAH lines -- somewhat lower than the ratio we observe, but with a
range that overlaps the 1$\sigma$ uncertainty for this galaxy.

\begin{figure*}
\plottwo{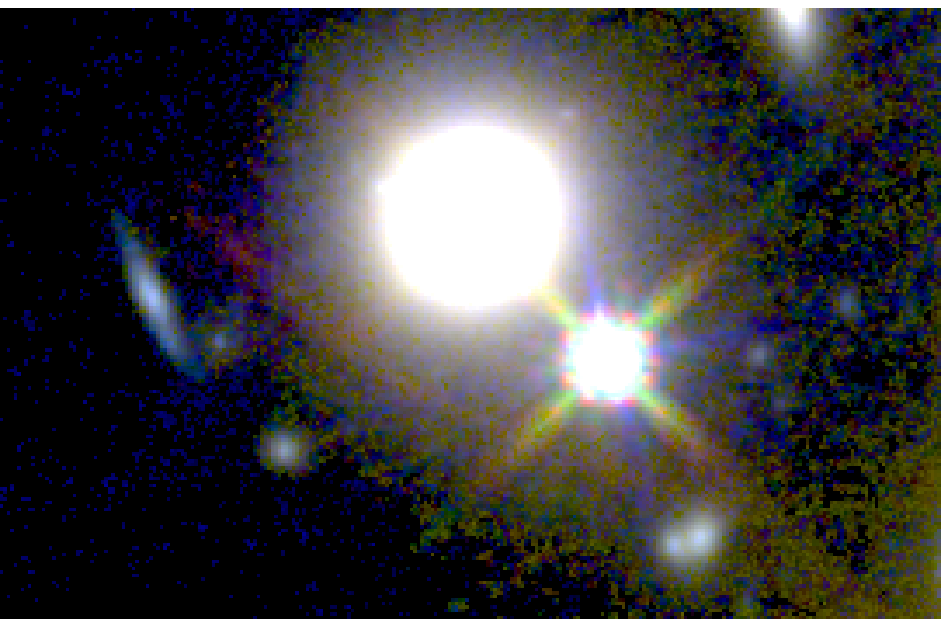}{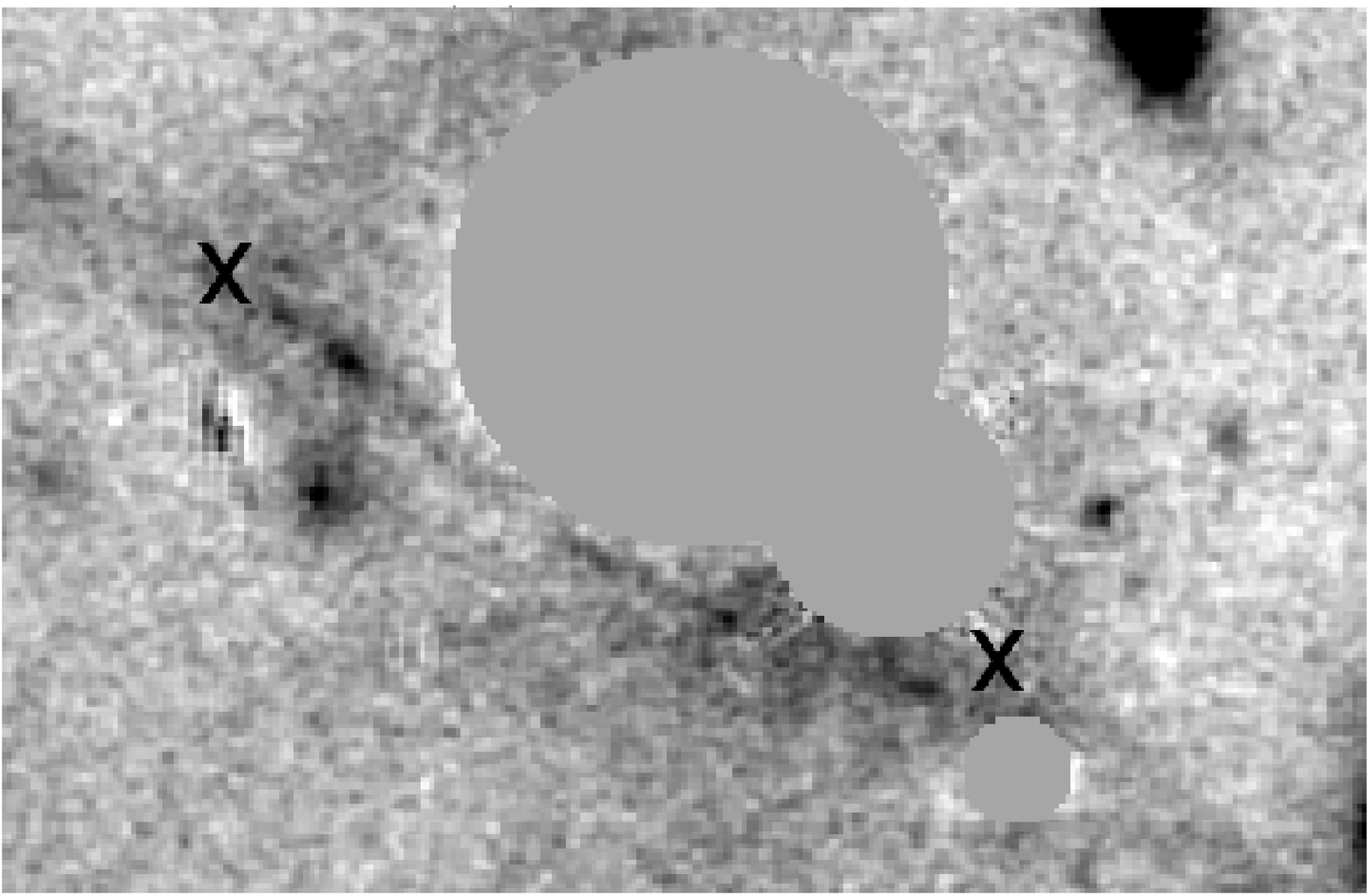}
\caption{\hst\ imaging of a $14\arcsec\times 9\arcsec$ region
  encompassing the lensed galaxy. The image on the left is a color
  composite using the F850LP, F110W, and F160W bands for blue, green,
  and red, respectively. Despite the bright foreground galaxy, the
  lensed galaxy can be seen as a red arc with two lobes. On the right
  we show the F160W image of the same region after the bright
  foreground objects have been modelled with \galfit\ and
  subtracted. In this image we also mask the cores of the three
  brightest objects for clarity. The arc, which is clearly visible in
  this image, breaks up into two lobes corresponding to images A and B
  in the \irac\ imaging from \firstp. Crosses denote the positions
  previously reported based upon \irac\ photometry.  The offsets are
  consistent with the uncertainties in the \irac\ positions due to
  source blending. North is up and east is to the
  left.\label{fig:arcims}}
\end{figure*}

\subsection{\wfc~Imaging of the Lensed Galaxy}
\label{subsec:wfc3}

With the new \hst~\wfc~imaging we repeat the photometric analysis
described in \firstp\ to try to identify the lensed galaxy in
high-resolution photometry.  We show in Figure \ref{fig:arcims} (left
panel) a composite color image constructed from the F850LP, F110W, and
F160 images.  The lensed galaxy can be seen in the composite image as
a faint red arc that contains brighter knots of emission in the
vicinity of images A and B from the \spitzer~data.

We use {\it GALFIT} version 3.0 to model and subtract foreground
objects near the location of the lensed galaxy. The object, which is
not detected in F850LP \firstpaper, is detected in F110W but is too
faint relative to the model residuals to recover robust photometry.
The arc is bright however in F160W (Figure \ref{fig:arcims}, right
panel). To verify that this arc is indeed the same lensed galaxy as
identified at longer wavelengths, we measure the flux ratio for the
two lobes of the arc. Specifically, we extract the fluxes within
polygonal apertures enclosing each of the bright emission regions and
correct for the sky using background apertures placed on nearby blank
sky regions. We measure a flux ratio $B/A=1.4\pm0.3$ -- consistent
with results at longer wavelengths in \firstp (see also \S
\ref{sec:mag}).  The total magnitude for the entire arc (images A and
B) is $m_{F160W}=23.80\pm0.1$ (AB), which corresponds to a flux
density of $1.1\pm0.1$ $\mu$Jy. As shall become clear in \S
\ref{sec:mag}, without magnification this object would be extremely
faint, with $m_{F160W}\approx28.8$.  We refrain from quoting a flux
density at F110W due to the lower contrast with the foreground galaxy
and higher associated systematic uncertainties.

\subsection{Magnification}
\label{sec:mag}

Conversion of the PAH line fluxes to luminosities requires not only
the redshift but also a determination of the lensing magnification. We
use the same mass maps described in \firstp.  Consequently, the only
changes from the previous analysis are the spectroscopic redshift and
improved astrometric positions. The former has only a modest impact,
whereas the latter have a more significant impact because of the steep
magnification gradient near the critical curve.

In the current discussion we are concerned primarily with images A and
B.  The \wfc~astrometry moves both images A and B slightly closer to
the critical curve and consequently raises the derived magnifications.
At the location of image A we compute a magnification $|\mu_A|\sim 32$
and at image B we obtain $|\mu_B|\sim69$. The measured flux ratio of
the two images from \firstp\ is $1.47\pm0.05$. To estimate the range
of plausible magnification for the blended \irs~spectrum, we take
individually the predicted magnifications for A and B at the locations
of peak emission for each arc and combine these with the observed flux
ratio.\footnote{Our magnification maps do not have the accuracy to
  resolve structure on the scale of the arcs, so we do use this point
  estimate rather than averaging over the photometric apertures.}  We
find that the total combined magnification is
$|\mu_{AB}|\sim80-115$. In subsequent discussion we will quote
quantities in terms of $\mu_{AB}/100$ to reflect the intrinsic
uncertainty in the magnification. In \firstp\ we used $\mu_A/25$,
which corresponds to $\mu_{AB}=80$.

\subsection{Presence of an AGN}

In \firstp\ we considered the question of whether the infrared
emission in this object is driven by star formation or AGN
activity. Based upon the hint of spatial extension in the \irac\ data
and lack of X-ray detection, we argued that the emission cannot be
purely due to AGN. We further concluded that the observed mid-infrared
colors (4.5 to 24 \micron) indicated that the source spectral energy
distribution may be a composite with contributions from both starburst
and AGN contributions.

We now revisit this question based upon both the \irs\ spectroscopy
and \wfc\ imaging.  From the spectral analysis we can also place an
upper bound on the contribution of the AGN to the mid-infrared
emission by estimating the fraction of the flux contained in the
power-law continuum.  From $5.7-8.8$ \micron\ rest-frame, the
continuum contributes $\sim 45$\% of the total emission, confirming
that the AGN is not dominant at mid-infrared wavelengths, but 
leaving the possibility that it may be a significant contributor to the far-infrared emission.

Additional information is provided by the \wfc\ imaging, in which we
resolve the target galaxy into gravitational arcs. As discussed in \S \label{subsec:wfc3},
the flux ratio for components A and B in the F160W observations is
consistent with the 24\micron\ flux ratio.  Because we are probing
near the critical curves where there are strong magnification
gradients, this should not necessarily be the case unless the stellar
emission and PAH emission have similar spatial distributions within
the galaxy. Thus, the \wfc\ data further supports the picture that in
this galaxy the mid- and far-infrared emission is dominated by star formation.

\begin{figure}
\plotone{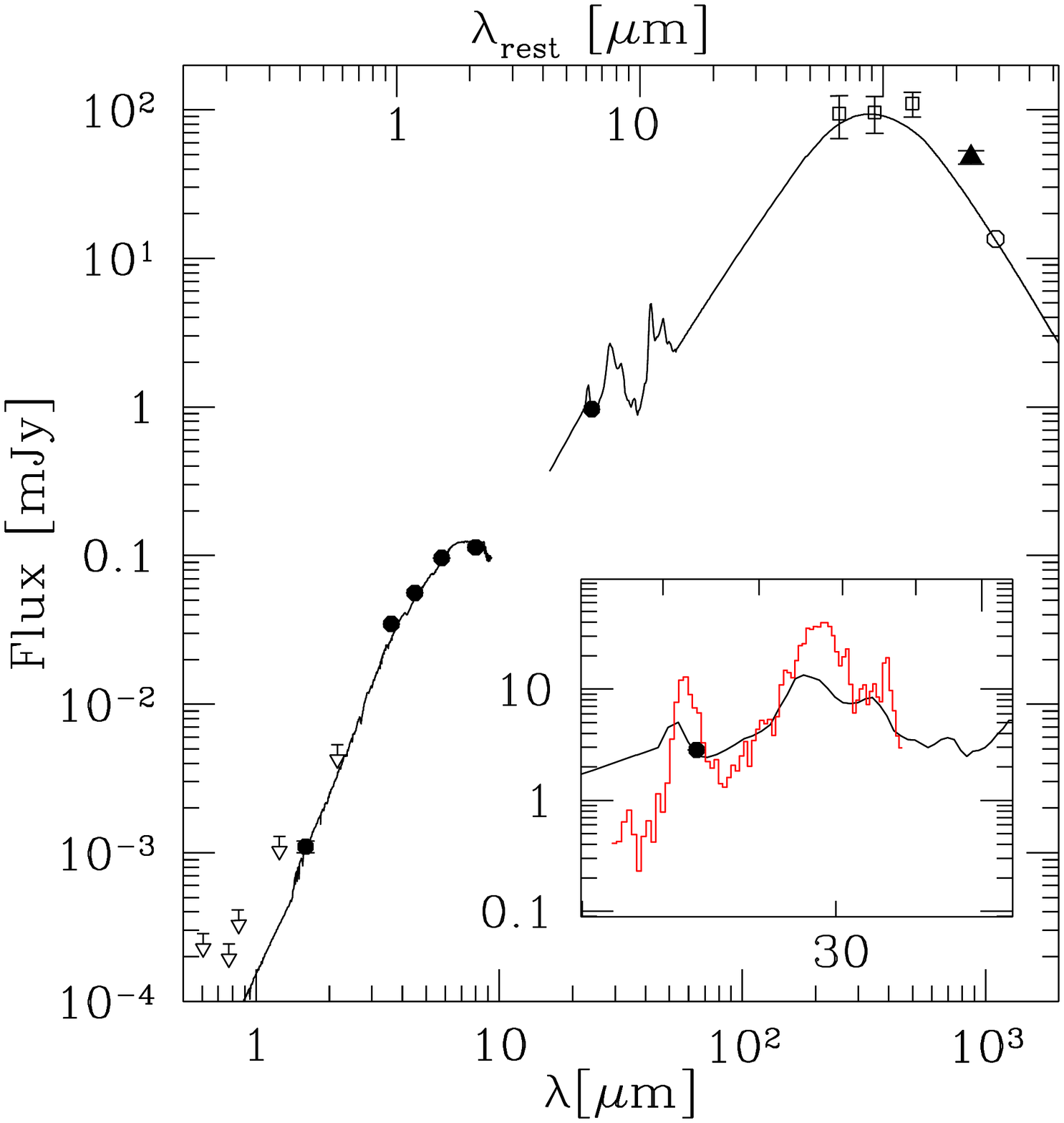}
\caption{Spectral energy distribution for the combination of images A and B (i.e. total flux density in A+B), including data points from \citet[open circle]{wilson2008}, \firstp, \citet[open squares]{blast2009}, and \citet[solid triangle]{johansson2010} in addition to the photometry and spectra presented in this paper. All downward pointing open arrows are upper limits. The solid lines corresponds to the best fit models from \hyperz\ and the \cite{chary2001} templates. The inset shows the model compared with the \irs\ spectrum (histogram) and 24\micron\ data point.
  \label{fig:sed}}
\end{figure}

\subsection{Stellar Population Age and Mass}
\label{sec:stellarmass}

We present in Figure \ref{fig:sed} the full spectral energy
distribution for the combination of images A and B, including all
published data.  Given the redshift, we can now repeat the analyses of
the SED presented in \firstp\ to refine the physical constraints on the
system.

We first use \hyperz~\citep{bolzonella2000} to derive improved
constraints on the age of the dominant stellar population and total
internal absorption of the system. As in \firstp, the input spectral
templates are based upon the Charlot \& Bruzual 2007 models with the
Padova 1994 evolutionary tracks \citep{bertelli1994} and a
\citet{chabrier2003} mass function. We use a \citet{calzetti2000}
extinction law to estimate the internal extinction, and fit the
photometry at observed wavelengths $\lambda<10\mu$m to exclude PAH emission.  We refer the
reader to \firstp\ for further details.  In deriving confidence
intervals on the age and extinction, we find that the data are best
fit by a 10 Myr population.
The solutions for the emission-weighted age bifurcate though,
permitting both very young templates and solutions with ages of a few
Gyrs. The 90\% confidence intervals on the age are $t<90$ Myrs and
$1.4<t<2.6$ Gyrs. We note however that strong observed submillimeter
emission provides a compelling argument for a very young stellar
population.  The extinction meanwhile is constrained to be
$A_V=3.8^{+0.5}_{-1.0}$ (90\% confidence). The confidence interval
reduces to $A_V=3.8^{+0.5}_{-0.2}$ if one considers only the younger
age solutions.

Next, we update our estimate of the stellar mass using the code
\kcorrect~\citep{blanton2007}, as in \firstp. We fit the combined flux
from images A and B, using the data at observed wavelengths of
1.5\micron\ $-$ 8\micron. For $z=2.79$ and $A_V=3.8$, we obtain a
stellar mass $M_*=4\times10^{9}$\invmab~$M_\sun$.\footnote{An
  important caveat to keep in mind is that systematic uncertainties in
  stellar mass estimates can yield a factor of two or greater errors
  \citep{conroy2009}.}  This value is roughly a factor of 3 lower than
the previously published result.  The added leverage from the
1.6\micron\ data point is the largest factor in the decrease, with the
addition of this data point accounting for roughly a factor of 2.  The
revised magnification estimates account for the remainder of the
change, while the changes in redshift and extinction have only a minor
impact.  The updated stellar mass implies that the target is a
low-mass dwarf galaxy.

\subsection{Total Infrared Luminosity}
\label{sec:lir}

There exist multiple different techniques for estimating the total
infrared luminosity for a galaxy where the infrared emission is
dominated by star formation.  The preferred technique is to use data
spanning the peak of the far-infrared (FIR) spectral energy
distribution and directly fit a modified blackbody to the far-infrared
emission.  This technique, which should now become the default method
in the era of \herschel, provides the most straightforward constraint
on $L_{IR}$ with the added advantage of directly yielding the
temperature of the dust if the redshift is known.  Many past studies,
in the absence of the requisite FIR data, have instead utilized either
24 \micron\ emission or PAH luminosities to estimate $L_{IR}$.  With the
unique data sets available for our lensed galaxy behind the Bullet
Cluster, we are able to compare estimates via all three approaches and
assess the level of consistency for this particular dusty starburst.

We first consider the direct approach.  \citet{blast2009} previously
reported $L_{IR}=2\times10^{11}$\invmab~L$_\sun$ and a dust
temperature $T\sim32$ K for this system using data from AzTEC and
BLAST with an assumed redshift of $z=2.9$.  \citet{johansson2010}
recently reported an 870 \micron\ flux density from the Large APEX
BOlometer CAmera (LABOCA) and noted that the 1.1 mm data point from
\citet{wilson2008} will soon be revised upward to $\sim20$ mJy. The
primary impact will be a $\sim10$\% revision in the dust temperature
and $L_{IR}$, which are both within the presumed uncertainties.  The
\herschel~Lensing Survey (PI: Egami) should soon significantly improve
determinations of both quantities \citep[see][]{rex2010}.

Second, we estimate $L_{IR}$ using the 24 \micron\ emission alone,
without separating the PAH from continuum emission.  This method is
predicated on matching the observed 24 \micron\ luminosity to a best-fit
spectral energy distribution. As in \firstp\ we use the code and
templates from \citet{chary2001} to calculate the bolometric infrared
luminosity. Updating the magnification and redshift from the previous
analysis, we obtain $L_{IR}=3\pm0.3\times10^{11}$
\invmab~L$_\sun$. This value is broadly consistent with the modified
blackbody fit to $24 \micron\ - 1$mm data, predicting well the shorter
wavelength BLAST data points and the published AzTEC 1.1 mm
result. The disagreement with the longer wavelength data points and
expected upward revision to the AzTEC data can be considered a
possible indication of a cooler dust temperature than in the
\citet{chary2001} templates.\footnote{A change in the normalization of the blackbody to match the BLAST and revised AzTEC data would yield a poor fit to the data near the peak; only a cooler dust temperature can simultaneously fit all the millimeter and submillimeter data.}

Finally, we consider the observed emission from the PAH features,
which are known to be strongly correlated with star
formation. \citet{pope2008} presented PAH luminosities for a sample of
13 SMGs (predominantly ultraluminous infrared galaxies with
$L_{IR}>2\times 10^{12}$) at $z=0.9-2.3$. Combining these with total
infrared luminosities from \citet{pope2006}, and including a sample of
lower luminosity low-redshift starburst galaxies, the authors derived
a relation between PAH and total infrared luminosity for
star-formation dominated infrared galaxies.

For the magnification derived in \S \ref{sec:mag}, we find that the
observed PAH features have luminosities $L(6.2\mu\mathrm{m}) = 2.4\pm
0.4 \times 10^{9}$ \invmab~L$_\sun$ and $L(7.7\mu\mathrm{m})=1.1\pm
0.2 \times 10^{10}$ \invmab~L$_\sun$.  Following the \citet{pope2008}
relations, we compute that $L_{IR}= 9.3\pm 1.6\times 10^{11}$
\invmab~L$_\sun$ and $L_{IR}=1.46\pm 0.27 \times10^{12}$
\invmab~L$_\sun$ from the 6.2 \micron\ and 7.7 \micron\ lines,
respectively. The quoted errors include the uncertainties in the PAH
measurements and in the coefficients of the \citet{pope2008}
$L_{PAH}-L_{IR}$ relations, but do not include intrinsic scatter in
the $L_{PAH}-L_{IR}$ relations.  The two PAH lines give values that
are factors of $5-7$ and $3-5$ higher than the estimates based upon
the FIR emission and 24 \micron\ template fitting, respectively.  This
offset can plausibly be argued as due to the intrinsic
scatter about the mean relation since the few galaxies in this
luminosity regime in \citet{pope2008} also exhibit large scatter;
however, similar offsets are also seen for the lensed
submillimeter galaxy SMM J1643554.2+661225, 
MS1512-cB58, and the Cosmic Eye, which all have $10^{11}$ M$_\sun <L_{IR}<10^{12}$ M$_\sun$ \citep{papovich2009,siana2009}. 
Moreover, \citet{rigby2008} observe a comparable enhancement in $L(8\mu\mathrm{m})/L_{IR}$ for LIRGs at $z\sim2$. Together these
results argue for a possible systematic shift relative to the \citet{pope2008}
relation at LIRG luminosities.

\subsection{Star Formation Rate}
\label{sec:sfr}
To estimate the star formation rate (SFR), we use the $L_{IR}$
determinations based upon the FIR emission and 24 \micron\ template
fitting. Following the \citet{kennicutt1998} relation, these $L_{IR}$
estimates imply SFR$\approx 100-150$ M$_\sun$ yr$^{-1}$.

Combining these values with the revised stellar mass, the estimated
specific star formation rate (SSFR) for this galaxy is $SSFR\approx$
25-40 Gyr$^{-1}$. The specific star formation rate and stellar mass
are consistent with the results for low-mass BM/BX and LBG galaxies in
\citet{reddy2006}. These results thus imply that this particular
galaxy is not unique, but rather indicative of the general low-mass,
star-forming population at this epoch.

\subsection{Molecular Hydrogen}
\label{sec:molecular}

We detect the $H_2\;0-0\;S(4)$ pure rotational line at rest frame
8.025 \micron\ in the \irs\ spectrum, and also have a tentative
($2\sigma$) detection of the $H_2\;0-0\;S(5)$ line at 6.909 \micron\
-- the first detection of these lines in a high redshift
galaxy. Fitting Drude profiles, we derive line fluxes of $f(H_2\;
S(4))=5.8\pm1.9\times10^{-15}$ \ergflux\ and $f(H_2\;
S(5))=2.5\pm1.2\times10^{-15}$ \ergflux\ (Table
\ref{tab:observed}). These fluxes correspond to luminosities of
$4.0\pm1.3\times10^{42}$ \invmab~erg s$^{-1}$ and
$1.7\pm0.8\times10^{42}$ \invmab~erg s$^{-1}$, respectively.  The
$S(4)$ line strength is exceptionally strong, with a
luminosity that exceeds that of the nearby LIRG NGC 6240
($L_{IR}\sim 7\times10^{10}$ L$_\sun$) by more than a
factor of 50 \citep{armus2006}.

The ratio of these two lines can be used to directly constrain the
excitation temperature of the gas.  Physically, for lower excitation
temperatures a greater fraction of the gas is in the low energy states
and the fraction of $H_2$ emitting in $S(4)$ decreases correspondingly.
  The observed ratio $f(H_2\;S(5))/f(H_2\;S(4))= 0.42\pm0.24$ implies
$T=377^{+68}_{-85}$ K for a single temperature gas -- relatively cool
compared to systems with shocked gas
\citep[e.g.][]{armus2006,egami2006}.  The corresponding constraint on
the warm molecular gas mass can then be computed as
\begin{equation}
M_{H_2} = m_H \frac{N(v,i)}{g_i} \mathrm{e}^{E(v,i)/kT} \sum_{j=0}^{\infty} g_j \mathrm{e}^{-E(v,j)/kT},
\end{equation}
where the $g_i$ are the statistical weights of the energy states \citep[see][]{rosenthal2000} and the number
density for the $S(4)$ transition can be calculated as 
\begin{equation}
N(v,i)=\frac{L(v,i)}{A(v,i)E(v,i)},
\end{equation}
where $L(v,i)$ is the 
luminosity and $A(v,i)$ is the Einstein coefficient. 
From this approach we derive a warm molecular gas mass 
$M_{H_2}=2.2^{+17}_{-0.8}\times10^8$
\invmab~M$_\sun$. This value serves as a lower bound on the total molecular gas
content of this system since we do not presently probe the cold gas. 
 Equivalently, this warm gas mass is $6^{+36}_{-4}$\% of
the stellar mass derived in \S \ref{sec:stellarmass}.  For comparison,
in their study of the lensed Lyman break galaxy LBG J213512.73-010143
\citet{coppin2007} use CO data to derive a cold molecular gas mass of
$2.4\times10^{9}$, which is 30\% of the stellar mass -- consistent
with what we see in this system (albeit with large uncertainty).
\begin{deluxetable}{cc}
\tabletypesize{\scriptsize}
\tablecaption{Summary Table of Derived Physical Properties}
\tablewidth{0pt}
\tablehead{
\colhead{Quantity} & \colhead{Value}
}
\startdata
z	      & $2.791\pm0.007$ \\
$|\mu_{AB}|$ & $80-115$ \\
$A_V$\tablenotemark{a} & $3.8^{+0.5}_{-1.0}$\\
Age\tablenotemark{a,b}& $<90$ Myrs\\
\\
SFR & 100-150 M$_\sun$ yr$^{-1}$\\
SSFR & 25-40 Gyr$^{-1}$ \\
M$^*$ & $4\times10^9$ M$_\sun$ \\
M$_{H_2}$ & $>4\times10^{8}$ M$_\sun$ \\
T$_{H_2}$ & $275-375$ K \\
& \\
L$_{IR}$ (24 $\mu$m) &$3\pm0.3\times10^{11}$ L$_\sun$  \\
L$_{IR}$ (FIR)\tablenotemark{c}& $2\times10^{11}$ L$_\sun$\\
L$_{IR}$ (PAH,6.2\micron) &  $9.3\pm1.6\times10^{11}$ L$_\sun$ \\
L$_{IR}$ (PAH,7.7\micron) &  $1.46\pm0.27\times10^{12}$ L$_\sun$ 
\enddata
\label{tab:physical}
\tablenotetext{a}{Confidence interval is 90\% rather than 1$\sigma$.}
\tablenotetext{b}{All quantities listed after age presume $\mu_{AB}=100$.}
\tablenotetext{c}{Value from \citet{blast2009}.}
\end{deluxetable}

The observed molecular emission is thus consistent with the picture of
a relatively cool ($T\sim375$ K) dusty starburst with a significant
gas reservoir.  A more detailed confirmation of this picture will
require future observations of either the $S(1) - S(3)$ lines or CO
rotational lines at longer wavelength, which will enable a robust
determination of the total gas content.  Finally, we note that the
lensed Lyman break galaxies J213512-01043 and MS1512-cB58 are the only
galaxies with similar $L_{IR}$ and redshift for which gas masses have
been determined via the CO (3-2) line
\citep{baker2004,coppin2007}.\footnote{The SMG SMM J2135-0102, which
  has a factor of $\sim4$ higher $L_{IR}$, also has a gas mass via the
  CO (1-0) line \citep{swinbank2010}.} The lensed galaxy described in
this paper exhibits much greater extinction and thus provides
complementary information.

\section{Summary and Discussion}
\label{sec:conclusions}

In this paper we present \irs\ spectroscopic confirmation of a
luminous infrared galaxy at $z=2.79$ that is highly magnified by the
Bullet Cluster. We also spatially resolve the galaxy with \wfc,
revealing gravitational arcs near the previously determined \spitzer\
positions.  The redshift determination and high signal-to-noise of the
spectrum, coupled with the multiwavelength photometry, enable us to
refine our previous analysis of the physical properties of this
galaxy. We determine that the lensed object is a low-mass dwarf
galaxy, M$_*=4\times10^{9}$ \invmab M$_\sun$, for which the
far-infrared thermal emission is star-formation dominated. The
inferred specific star formation rate, $SSFR\approx 25-40$ Gyr$^{-1}$,
is typical for a field galaxy of this mass at this redshift
\citep[e.g.][]{reddy2006}. The observed and derived physical
quantities are summarized in Tables \ref{tab:observed} and
\ref{tab:physical}.

By virtue of the extreme lensing -- the two brighest images have a
combined magnification of roughly 100 -- this object thus provides a
powerful laboratory in which to probe the conditions in low-mass,
star-forming galaxies during an epoch when they are otherwise
inaccessible. Indeed, this galaxy is 
one of the two lowest mass systems
with \irs\ spectroscopy at $z>2$.\footnote{The lensed LBG MS1512-cB58 has a factor of $\sim2-4$ lower stellar mass \citep[see][]{siana2008}.} Perhaps most promising 
for future investigations, our detection of
rotational H$_2$ lines is indicative of a large molecular gas
reservoir. We derive a temperature of $T=377^{+68}_{-85}$ K and
estimate a gas mass of $M_{H_2}=2.2^{+17}_{-0.8}\times10^8$~M$_\sun$
in this component, roughly 2-42\% of the stellar mass.  Future
spectroscopic observations of longer wavelength H$_2$ lines with
\herschel\ and CO rotational lines with a facility such as the Large
Millimeter Telescope or the Atacama Pathfinder EXperiment (APEX)
therefore have the potential to precisely measure the total molecular gas
content in this galaxy.  Finally, we note that because
dwarf galaxies of this mass are a ubiquitous population, the odds are
good that programs like the \herschel\ Lensing Survey can detect a
sizable sample of similar lensed sources behind massive clusters.

\acknowledgements 
AHG and CP thank Eiichi Egami, Jane Rigby, and Ranga-Ram Chary for 
constructive discussions related to this work. The authors also thank
the anonymous referee for comments that improved this paper.
This work is based on observations made with the Spitzer Space
Telescope, which is operated by the Jet Propulsion Laboratory,
California Institute of Technology under a contract with NASA. Support
for this work was provided by NASA through an award issued by
JPL/Caltech.  The authors acknowledge support for this work from
NASA/HST grants HST-GO-10200, HST-GO-10863, and HST-GO-11099, as well
as NASA/Spitzer grants 1319141 and 1376614.

{\it Facilities:} HST (ACS,WFC3), Spitzer (IRAC,IRS,MIPS), Magellan:Baade (PANIC) 
\bibliographystyle{apj} \bibliography{ms}

\end{document}